\documentclass[12pt,preprint]{aastex}
\usepackage{amssymb}
\usepackage{epsfig}
\usepackage{natbib}
\usepackage{graphicx}
\usepackage{color}
\usepackage{tabularx}
\usepackage{epsfig}
\usepackage{rotating}
\usepackage{multirow}
\usepackage{subfigure}
\usepackage{ulem}
\usepackage{longtable}

\newcommand       \Ks           {{K_{\rm s}}}

\shorttitle{Twelve-Band Period--Luminosity Relations for W UMa-type Contact Binaries}
\shortauthors{X. Chen et al.}

\begin{document}

\title{Optical--Mid-Infrared Period--Luminosity Relations for W
  UMa-type Contact Binaries Based on Gaia DR 1: 8\% Distance Accuracy}

\author{Xiaodian Chen\altaffilmark{1}, 
Licai Deng\altaffilmark{1}, 
Richard de Grijs\altaffilmark{2,3,4},
Shu Wang\altaffilmark{3}, 
and Yuting Feng\altaffilmark{5}
}
\altaffiltext{1}{Key Laboratory for Optical Astronomy, National
  Astronomical Observatories, Chinese Academy of Sciences, 20A Datun
  Road, Chaoyang District, Beijing 100012, China;
  chenxiaodian@nao.cas.cn}
\altaffiltext{2}{Department of Physics and Astronomy, Macquarie
  University, Balaclava Road, North Ryde NSW 2109, Australia;}
\altaffiltext{3}{Kavli Institute for Astronomy \& Astrophysics, Peking
  University, Yi He Yuan Lu 5, Hai Dian District, Beijing 100871,
  China;}
\altaffiltext{4}{International Space Science Institute--Beijing, 1
  Nanertiao, Zhongguancun, Hai Dian District, Beijing 100190, China;}
\altaffiltext{5}{Department of Astronomy, Peking University, Yi He
  Yuan Lu 5, Hai Dian District, Beijing 100871, China}

\begin{abstract}
W Ursa Majoris (W UMa)-type contact binary systems (CBs) are useful
statistical distance indicators because of their large numbers. Here,
we establish (orbital) period--luminosity relations (PLRs) in 12
optical-to-mid-infrared bands ($GBVRIJH\Ks W1W2W3W4$) based on 183
nearby W UMa-type CBs with accurate Tycho--{\sl Gaia} parallaxes. The
1$\sigma$ dispersion of the PLRs decreases from optical to near- and
mid-infrared wavelengths. The minimum scatter, 0.16 mag, implies that
W UMa-type CBs can be used to recover distances to 7\%
precision. Applying our newly determined PLRs to 19 open clusters
containing W UMa-type CBs demonstrates that the PLR and open cluster
CB distance scales are mutually consistent to within 1\%. Adopting our
PLRs as secondary distance indicators, we compiled a catalog of 55,603
CBs candidates, of which 80\% have distance estimates based on a
combination of optical, near-, and mid-infrared photometry. Using
Fourier decomposition, 27,318 high-probability W UMa-type CBs were
selected. The resulting 8\% distance accuracy implies that our sample
encompasses the largest number of objects with accurate distances
within a local volume with a radius of 3 kpc available to date. The
distribution of W UMa-type CBs in the Galaxy suggests that in
different environments, the CB luminosity function may be different:
larger numbers of brighter (longer-period) W UMa-type CBs are found in
younger environments.
\end{abstract}

\section{Introduction}

W Ursa Majoris (W UMa)-type contact binaries (CBs) represent a class
of short-period eclipsing binary systems. Unlike detached and
semi-detached eclipsing binaries, the two components of W UMa-type CBs
fill their Roche lobes. W UMa-type CBs are fainter and have shorter
periods than early-type CBs. The radii of both components are
correlated with their orbital periods. Based on Roche lobe theory
\citep{Eggleton83}, the mean density is related to the orbital period,
which can, in turn, be used to derive the period--luminosity--color
relation (PLCR). \citet{Rucinski94} established the first accurate
observational PLCR, eventually reaching a precision of 12\%
\citep{Rucinski97}. Using near-infrared data, which are less sensitive
to temperature, extinction, and metallicity variations than optical
observations, \citet{Chen16b} established the first accurate CB
(orbital) period--luminosity relations (PLRs) based on 66 CBs. These
PLRs have 10\% (1$\sigma$) accuracy, which renders W UMa-type CBs a
potential reliable statistical distance tracer. Characterized by
similar properties as W UMa-type CBs, early-type CBs \citep{Chen16b}
and red giant CBs \citep{Muraveva14} also obey PLRs, although the
associated scatter is larger. Taking advantage of Tycho--{\sl Gaia}
astrometric solution (TGAS) parallaxes, studies based on W UMa-type
CBs can potentially benefit significantly because of the large number
of such objects in the solar neighborhood. Applying these parallaxes,
\citet{Mateo17} found a ($\log P,M_V$) PLR slope of $-9.0\pm0.4$,
which is in good agreement with the slope of $-9.15\pm0.12$ from
\citet{Chen16b}. However, some open questions still remain, such as
the limiting distance precision and any applicable zeropoint
constraints.

W UMa-type CBs are one of the most numerous variables in the Galaxy,
since they cover the evolutionary phases of binaries with long orbital
timescales. The General Catalog of Variable Stars
\citep[GCVS][]{Samus17} lists 1131 CBs, while the All Sky Automated
Survey (ASAS) Catalog of Variable Stars \citep{Pojmanski05} includes
5374 CBs candidates. The largest CB sample available to date is based
on the Catalina Survey and Catalina Survey southern catalogs,
containing 30,710 and 18,803 CB candidates, respectively. In addition,
hundreds of individual W UMa-type CBs are detected every year, so that
a catalog of high-probability CBs with distance information can
readily be constructed. The associated distance scale will benefit not
only from constraining the absolute physical parameters of CBs, but
also from the reduced uncertainties associated with the anticipated
{\sl Gaia} Data Release (DR) 2 parallaxes.

In this paper, we aim at establishing PLRs for W UMa-type CB samples
covering wavelengths from the optical to the near- and
mid-infrared. We will evaluate the optimal accuracy of W UMa-type CBs
as distance tracers. A complete catalog of CB candidates is compiled
and strict criteria to distinguish CBs from other types of variables
are presented. The spatial distribution of CBs in our Galaxy is also
discussed. In Section 2, we present the adopted data sets and our
catalog. The steps taken to establish the 12-band PLRs are discussed
in Section 3. The distances to open clusters based on W UMa-type CBs
are estimated and compared in Section 4. A catalog of high-probability
CBs with distance information is discussed in Section 5. We summarize
our main conclusions in Section 6.

\section{Data}
\subsection{W UMa-type contact binaries with TGAS parallaxes}

For our CB sample, distance information is based on their TGAS
parallaxes. TGAS contains two million stars with parallaxes
\citep{Gaia16a, Gaia16b}, which includes all nearby CBs. The average
random parallax uncertainty for the entire sample is 0.3 mas; even for
the nearest CBs, the random uncertainty is still greater than 0.2
mas. In addition, the TGAS parallaxes are known to be affected by a
{\it systematic} error of 0.3 mas, which cannot be reduced when
dealing with small regions on the sky \citep[see also][their Section
  3]{Gaia16b}. Since CBs are distributed randomly across the full sky,
this systematic error is equivalent to the random uncertainty for the
whole sample. The uncertainties pertaining to the CB PLRs are mainly
owing to propagation of the uncertainties associated with the parallax
distances. To achieve high-accuracy PLRs, CBs with larger parallax
uncertainties should consequently be eliminated. To ensure both
sufficient numbers and small parallax uncertainties in our final CB
sample, we selected objects within 330 pc ($\pi>3$ mas) with parallax
uncertainties of better than 7\% (0.16 mag), resulting in a total
sample size of 204. If we adopt a maximum systematic parallax error of
0.3 mas, the average total uncertainty in the distance modulus is
$0.16\pm0.05$ mag for this sample.

The photometric data we have access to consist of $(B-V)$ colors from
the Tycho catalog, maximum $V$-band magnitudes ($V_{\rm max}$) from
the ASAS catalog, $G_{\rm mean}$ magnitudes from {\sl Gaia} DR 1, $RI$
magnitudes from the U.S. Naval Observatory (USNO)-B catalog
\citep{Monet03}, near-infrared $JH\Ks$ data from the Two Micron
All-Sky Survey (2MASS) catalog \citep{Cutri03}, and $W1, W2, W3$, and
$W4$ magnitudes from the Wide-Field Infrared Survey Explorer (WISE)
catalog \citep{Wright10}. The ASAS catalog covers most of the sky,
with $V$-band CB light curves in the range $8<V<14$ mag. For each
object, more than one hundred detections are recorded, so it is easy
to determine both maximum and average magnitudes. {\sl Gaia} DR 1
includes ten billion objects with $G$-band mean magnitudes based on
more than one hundred detections each. The $G$ band is centered at 673
nm and the filter has a width of 440 nm \citep{Jordi10}. 2MASS
provides single-epoch $JH\Ks$ photometry but it includes the
observations' Julian dates, which can be converted to maximum
magnitudes using template light curves. The details pertaining to our
conversion from single epochs or mean magnitudes to the corresponding
maximum magnitudes are discussed in Section 3.1.  WISE is a full-sky
survey undertaken in four mid-infrared bands: W1 (3.35 $\mu$m), W2
(4.60 $\mu$m), W3 (11.56 $\mu$m), and W4 (22.09 $\mu$m). WISE has
20--100 detections for each object, which means that the WISE catalog
is a powerful tool to find and study variable stars. The limiting
magnitudes in the four bands are 16.5, 15.5, 11.2, and 7.9 mag,
respectively; except for the $W4$ band, all 194 W UMa-type CBs have
sufficient signal-to-noise ratios for our analysis.

\subsection{The catalog of contact binaries}

The GCVS contains 1131 CBs, which were collected from individual
studies. Those CBs represent a genuine CB sample, because most of
these CBs are well observed and carefully studied. The ASAS catalog
lists 5374 CBs. This sample only includes a small number of
incorrectly identified CBs, since the authors separated CBs from
semi-detached binaries. In addition, they only collected about 100 CBs
with amplitudes of less than 0.15 mag \citep{Rucinski06} to avoid
sample contamination. In recent years, the Catalina \citep{Drake14}
and Catalina southern \citep{Drake17} catalogs have yielded about
50,000 CB candidates covering the declination range
$-70^\circ<\delta<65^\circ$ but avoiding the Galactic plane. These
latter catalogs reach down to at least $V=19$ mag. Unlike the ASAS
catalog, the Catalina catalogs do not distinguish between CBs and
semi-detached binaries and they also contain a number of low-amplitude
CBs. Therefore, these CB candidates need further study.

The total number of CB candidates in our sample is 56,603 and the
photometric data adopted cover the {\sl Gaia} $G$ band, the
near-infrared $JH\Ks$ bands, and the mid-infrared $W1$ filters, since
they are homogeneous and deep enough.

\begin{table}[h!]
\begin{center}
\caption{\label{t1} Twelve-band magnitudes for 183 W UMa-type CBs with TGAS parallaxes.\tablenotemark{a}}
\tiny
\vspace{0.1in}
\begin{tabular}{rrcccccccccccc}
\hline \hline
R.A. (J2000) &  Dec. (J2000)& Period & ${\rm DM}_{\rm TGAS}$ &${\rm err_{DM}}$ & Amp. &$A_V$& $M_{V_{\rm max}}$ & $ M_{G_{\rm max}}$ & $(B-V)$ & $M_{R_{\rm max}}$ & $M_{I_{\rm max}}$ & $M_{J_{\rm max}}$  & ... \\
 ($^\circ$) & ($^\circ$)    & [d] & (mag) & (mag) & (mag) & (mag) & (mag)& (mag)  & (mag) & (mag) & (mag) & (mag)  & ... \\
\hline
        27.80248  &  43.81867  &   0.38303  &  6.868  &  0.118  &  0.53 &0.10 &  4.362  &  4.043  &  0.823  &  3.728  &  3.408  &  2.959   &...\\
    31.83356  &  35.64864  &   0.38970  &  6.763  &  0.147  &  0.45 &0.09 &  4.087  &  3.905  &  1.134  &  3.432  &  3.002  &  2.686   &...\\
    42.17014  &  13.74623  &   0.28234  &  5.493  &  0.065  &  0.27 &0.05  &  4.787  &  4.917  &  0.917  &  4.707  &  4.357  &  3.954   &...\\
... &...&...& ... & ... & ... & ... & ... & ... & ... & ... & ...& ...& ...\\
\end{tabular}
\tablenotetext{a}{${\rm DM}_{\rm TGAS}$ and ${\rm err_{DM}}$ denote
  the distance modulus and its uncertainty derived from TGAS
  parallaxes; Amp. is the amplitude of the $V$-band light curve and
  $A_V$ is the $V$-band extinction. $M_{\lambda_{\rm max}}$ are the
  maximum absolute magnitudes in each band, while $M_{\lambda_{\rm
      mean}}$ are the mean absolute magnitudes. The entire table is
  available in the online journal. A portion is shown here for
  guidance regarding its form and content.}
\end{center}
\normalsize
\end{table}

\section{Twelve-band PLRs for W UMa-type contact binaries}
\citet{Chen16b} first determined the near-infrared $JH\Ks$ CB PLRs
based on 66 CBs with open cluster distances and {\sl Hipparcos}
parallaxes. In this paper, we aim at making the PLRs of W UMa-type CBs
more complete, accurate, and convenient for follow-up use. First,
photometric observations in the $G$ band and mid-infrared bands are
introduced, since numerous detections are available in these
filters. Johnson $BVRI$ magnitudes are added because they are widely
used. Second, \citet{Chen16b} only used maximum absolute magnitudes to
establish their PLRs. In contrast, here we intend to establish PLRs
based on both maximum and mean magnitudes. Although in theory only the
maximum magnitude is related to the orbital period, a relation between
the mean magnitude and the orbital period is obvious, because the
dispersion between the maximum and mean magnitudes is only
$\sigma=0.05$ mag (see Section 3.1). However the mean-magnitude PLR is
more suitable for use with large-sample surveys with sparse period
coverage, since mean magnitudes are easier and better determined than
maximum magnitudes. Third, only W UMa-type CBs are considered, since
the number of early-type CBs in our sample (16) is too small for
further analysis.

\subsection{Light curve analysis}

To establish multi-band PLRs, the most important problem we must
overcome is the conversion from magnitude in a single epoch to the
corresponding maximum magnitude. To address this problem, we first
need to know how the light curve shape changes with
wavelength. Luminosity variations of CBs are caused by geometric
eclipses of the two binary components. During an eclipse, the
temperature change is very small, since the temperatures of both
components are similar. Therefore, almost no light curve shape
variation occurs between different bands. This is contrary to the
situation for pulsating stars, whose amplitude decreases with
increasing wavelength. To make sure that this situation indeed applies
to our CBs, we compared the $W1$ and $V$-band light curves for the 204
calibration CBs and found that they are in good agreement. Since the
light curve shape does not change with wavelength, we can use the
$V$-band light curve as our template light curve, and 2MASS magnitudes
obtained at a given epoch can hence be converted to maximum
magnitudes.

The second issue of importance is the conversion between mean and
maximum magnitudes. Statistically speaking, the difference between
both magnitudes is proportional to the light curve amplitude:
$\Delta=(0.361\pm0.001) {\rm Amp}+(0.016\pm0.000)$, $\sigma=0.017$ mag
for the 27,000 genuine CBs in the Catalina catalog. Based on this
equation, {\sl Gaia} $G_{\rm mean}$ magnitudes can be converted to
$G_{\rm max}$ magnitudes. If the amplitude is not taken into account,
the difference between the mean and maximum magnitudes is
$\Delta_1=0.14\pm0.05$ mag; the 0.05 mag dispersion will propagate to
the scatter in the mean-magnitude PLR. For the USNO-B1, single-epoch
$RI$ magnitudes are used to derive the PLRs, because the observation
time are not recorded. This will introduce a statistical error of
$\Delta_1/2 = 0.07$ mag and a possible systematic bias of
$\Delta_1/\sqrt {183} = 0.01$ mag if we assume that the 183 W UMa-type
CBs with accurate TGAS parallaxes (see below) are equally distributed
in phase. Since this bias is negligible and the statistical
uncertainty is small, these $RI$ PLRs are acceptable for inclusion in
our analysis.

\subsection{Period--Luminosity Relations}

\begin{figure}[h!]
\centering
\vspace{-0.0in}
\includegraphics[angle=0,width=160mm]{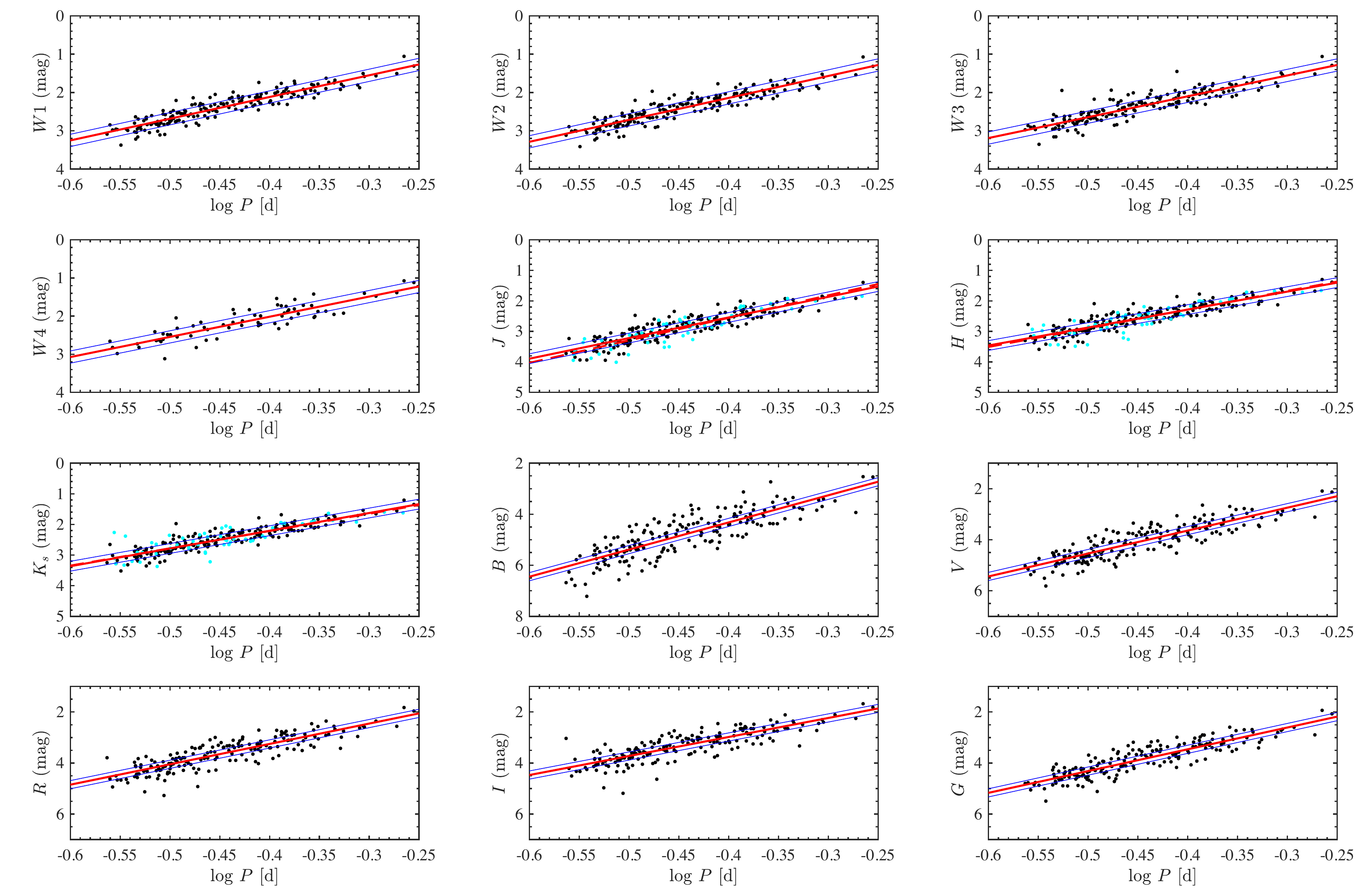}
\vspace{-0.0in}
\caption{\label{f1} Twelve-band PLRs for W UMa-type CBs using their
  maximum magnitudes. The black data points are the 183 W UMa-type
  CBs, while the red lines are the best fits. The blue lines represent
  the best fits with typical uncertainties of $\pm0.16$ mag. The cyan
  dots in the $J$, $H$, and $\Ks$ panels are W UMa-type CBs from
  \citet{Chen16b} and the red dashed lines are their best fits.}
\end{figure}

We selected 183 of the 204 CBs to determine the PLRs (see Table
\ref{t1}), excluding early-type CBs with long periods, $\log P>-0.25$
[days], since they are rare, and short-period CBs, $\log
P<-0.575$ [days], since the latter are fainter than the other
stars \citep{Chen16b, Mateo17}. The period range covered by our CB
sample spans approximately 0.3 days, which is similar to the range of
RR Lyrae \citep{Gaia17}. The maximum-magnitude PLRs are shown in
Fig. \ref{f1}. The absolute magnitudes were estimated based on the
equation $M_{\lambda}=m_{\lambda}-5{\rm log}(1000/\pi)+5-A_{\lambda}$,
where $\pi$ is the parallax in units of mas.

 Extinction values were estimated for each CB individually,
 i.e.,\\ $A_V=\rho_0 \int_{0}^{d} {\exp(-r|\sin(b)|/H) {\rm d}r}$
 \citep[c.f.][]{Nataf13, Mateo17}, where $b$ represents the object's
 Galactic latitude and $d$ is the distance derived from the
 corresponding TGAS parallax. $H = 164$ pc is the dust scale height
 \citep{Nataf13}. This small scale height indicates that the dust is
 predominantly concentrated in the Galactic plane. The extinction is
 estimated based on the length of sightline through the dust
 distribution, and hence it reflects global extinction variations with
 distance $d$ and Galactic latitude $b$. The mean density of the dust
 in the Galactic plane, $\rho_0=0.54$ mag pc$^{-1}$, was estimated
 based on 2000 nearby OCs \citep{Chen16b}, which span a similar
 (nearby) distance range as the CBs. Statistically speaking, the
 average extinction for our 183 CBs is $A_V=0.075\pm0.025$ mag. This
 is a reliable extinction value for an average distance of 183 pc. The
 extinction values in the other bands were estimated based on the
 relative extinction values $A_{\lambda}/A_V$ from \citet{Rieke85} and
 \citet{Jordi10}. The extinction model can be used to estimate
   the extinction values with 20\% accuracy. Therefore, after the
   correction, the remaining systematic bias introduced by extinction
   variations is around 0.015 mag in the $V$ band and 0.0015 mag in
   $W1$. This is significantly smaller than the uncertainties in the
   PLRs.

The red lines in Fig. \ref{f1} are the linear fits to the twelve-band
$M_\lambda$ versus $\log P$ trends. Their slopes and the intercepts
obtained from application of the least-squares method are listed in
Table \ref{t2} for both the maximum- and mean-magnitude PLRs. The
numbers used to determine the PLRs are also included. Only for $W4$,
those objects that exceed the detection limit are excluded. A
comparison of the maximum- and mean-magnitude PLRs shows that the
scatter in the mean-magnitude PLRs is somewhat larger, which is
accordance with the analysis in the previous section. We also adopted
the bootstrap-sampling technique to explore possible variations in the
coefficients and the PLR uncertainties. The best-fitting coefficients
and 1$\sigma$ uncertainties resulting from application of 10,000
bootstrap-sampling runs are identical to those based on the
least-squares method, except for the $W4$ band. In the latter case,
the PLRs resulting from the bootstrap-sampling technique are
$M_{W4}({\rm max})=(-5.28\pm0.32)\log P+(-0.10\pm0.14), \sigma=0.21$
mag and $M_{W4}({\rm mean})=(-5.55\pm0.33)\log P+(-0.05\pm0.14),
\sigma=0.21$ mag. The differences with respect to the values returned
by the least-squares method (Table \ref{t2}) are negligible.  Although maximum-magnitude PLRs are more accurate than
  mean-magnitude relations, mean-magnitude PLRs are more practical
  when dealing with large surveys or faint objects. For such data
  sets, light curves are usually sparsely covered and affected by
  large photometric uncertainties. Few or no good light curves are
  usually available to determine the maximum magnitudes.

\begin{figure}[h!]
\centering
\vspace{-0.0in}
\includegraphics[angle=0,width=160mm]{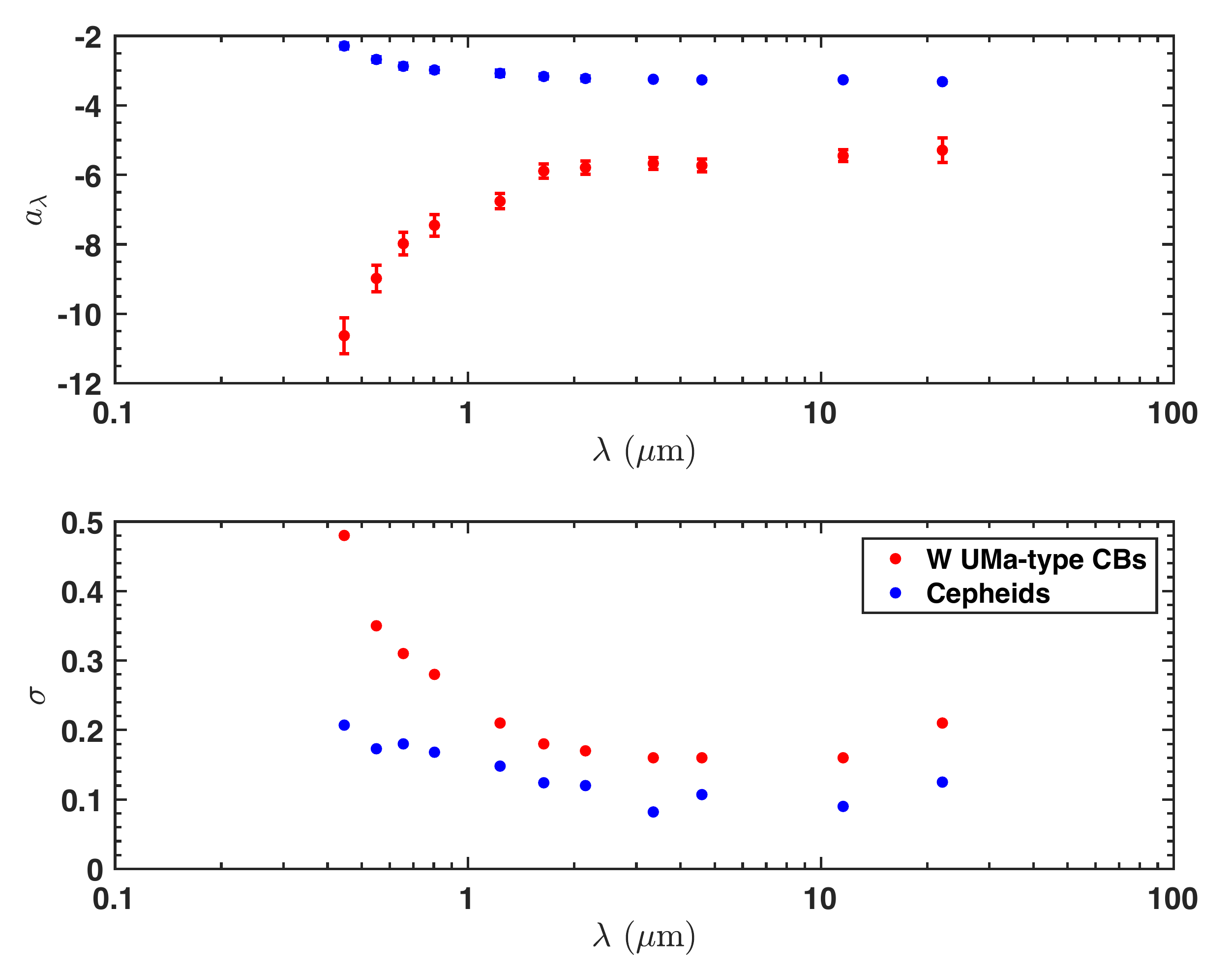}
\vspace{-0.0in}
\caption{\label{f2} (Top) The PLR slopes change as a function of
  wavelength for W UMa-type CBs (red) and Cepheids (blue), while the
  $1\sigma$ dispersion is shown in the bottom panel.}
\end{figure}

By studying the PLRs in different bands, an obvious trend is found in
the slopes and the scatter properties. The slopes increase with
increasing wavelength, while the scatter decreases. We compare this
behavior with that of the PLRs of classical Cepheids \citep{Fouque07,
  Chen17, Wang18} in Fig. \ref{f2}. In the top panel, the slopes of
the Cepheid PLRs decrease with wavelength, which is contrary to what
we saw for the W UMa-type CBs. The reason for this is that Cepheids
become redder when they increase in brightness, while W UMa-type CBs
become bluer. In addition, Cepheids cover a larger magnitude range at
long wavelengths, while W UMa-type CBs span a larger magnitude range
at short wavelengths. We also found that the two trends become flat
for wavelengths exceeding 2 $\mu$m ($\Ks$ band), which means that
mid-infrared colors are insensitive to temperature. This is also found
in the bottom panel: the $1 \sigma$ scatter first decreases with
increasing wavelength, but it becomes flat in the mid-infrared. As
regards this scatter, temperature is one contributor, but the other
contributors (such as extinction and metallicity variations) also
decrease with increasing wavelength. The photometric and distance
errors therefore do not depend on wavelength. This scatter represents
the precision of variables as distance indicators if the external
errors are well constrained. For Cepheids, the typical uncertainty is
0.10 mag (5\%) while for W UMa-type CBs it is 0.16 mag (7\%). Compared
with Cepheids, W UMa-type CBs are not as well studied, so reduction of
of this scatter is expected based on future studies.

\begin{table}[h!]
\begin{center}
\caption{\label{t2} Parameters of the Galactic mid-infrared PLRs}
\vspace{0.1in}
\begin{tabular}{ccccc}
\hline \hline
Filter ($\lambda$) & $N$ & $a_\lambda$  & $b_\lambda$ & $\sigma$\\
\hline
\multicolumn{5}{c}{$M_\lambda({\rm max})=a_\lambda\times\log P+b_\lambda$}\\
\hline
$W1$ &183 & $-5.67 \pm0.17$ & $-0.15\pm0.08$ & 0.16\\
$W2$ &183 & $-5.73 \pm0.18$ & $-0.15\pm0.08$ & 0.16\\
$W3$ &182 & $-5.45 \pm0.17$ & $-0.08\pm0.08$ & 0.16\\
$W4$ &67  & $-5.29 \pm0.35$ & $-0.10\pm0.15$ & 0.21\\
$G $ &183 & $-8.53 \pm0.33$ & $0.05 \pm0.15$ & 0.31\\
$B $ &183 & $-10.63\pm0.52$ & $0.07 \pm0.23$ & 0.48\\
$V $ &183 & $-8.98 \pm0.38$ & $0.05 \pm0.17$ & 0.35\\
$R $ &183 & $-7.98 \pm0.33$ & $0.06 \pm0.15$ & 0.31\\
$I $ &183 & $-7.45 \pm0.31$ & $0.00 \pm0.14$ & 0.28\\
$J $ &183 & $-6.76 \pm0.22$ & $-0.16\pm0.10$ & 0.21\\
$H $ &183 & $-5.89 \pm0.20$ & $-0.07\pm0.09$ & 0.18\\
$\Ks$&183 & $-5.79 \pm0.19$ & $-0.11\pm0.08$ & 0.17\\  

\hline
\multicolumn{5}{c}{$M_\lambda({\rm mean})=a_\lambda\times\log P+b_\lambda$}\\
\hline
$W1 $ &183 & $-5.83 \pm0.18$ & $-0.05\pm0.08$ &0.17\\
$W2 $ &183 & $-5.87 \pm0.19$ & $-0.04\pm0.09$ &0.18\\
$W3 $ &182 & $-5.66 \pm0.18$ & $0.01 \pm0.08$ &0.17\\
$W4 $ &67  & $-5.55 \pm0.36$ & $-0.04\pm0.16$ &0.22\\
$G  $ &183 & $-8.58 \pm0.35$ & $0.20 \pm0.16$ &0.32\\
$B  $ &183 & $-10.79\pm0.54$ & $0.17 \pm0.24$ &0.51\\
$V  $ &183 & $-9.14 \pm0.40$ & $0.16 \pm0.18$ &0.37\\
$R  $ &183 & $-8.06 \pm0.34$ & $0.19 \pm0.15$ &0.32\\
$I  $ &183 & $-7.64 \pm0.32$ & $0.09 \pm0.14$ &0.29\\
$J  $ &183 & $-6.87 \pm0.25$ & $-0.04\pm0.11$ &0.23\\
$H  $ &183 & $-5.99 \pm0.21$ & $0.05 \pm0.10$ &0.20\\
$\Ks$ &183 & $-5.95 \pm0.21$ & $0.00 \pm0.09$ &0.19\\

\hline
\hline
\end{tabular}
\end{center}
\end{table}

Focusing on the scatter in the W UMa-type CB PLRs, in the optical the
uncertainty is around 0.3 mag. To deal with this obvious dependence on
temperature in the scatter, a second band is needed to establish
PLCRs. A combination of the $G$ and $W1$ bands yields the smallest
uncertainty, $M_{G}({\rm max})=-5.19\log P+1.17(G-W1)_0-0.19,
\sigma=0.16$ mag (7\%) and $M_{G}({\rm mean})=-5.02\log
P+1.28(G-W1)_0-0.11, \sigma=0.16$ mag (7\%). For the $V$ band,
$M_{V}({\rm max})=-5.27\log P+1.12(V-W1)_0-0.18, \sigma=0.16$ mag
(7\%) and $M_{V}({\rm mean})=-5.20\log P+1.19(G-W1)_0-0.09,
\sigma=0.16$ mag (7\%). This also suggests that a 7\% distance
accuracy can be achieved using this combination of $G$ ($V$) and $W1$
magnitudes. Note that the parallax uncertainty is around 0.16 mag (see
Section 2.1), so the best accuracy attainable is limited by that in
the parallax distance. We are confident that the accuracy will benefit
significantly from {\sl Gaia}'s five-year parallaxes, once made
available.
\subsection{The PLR Zeropoint}

Except for the uncertainty, the zeropoint is the other parameter that
can be used to evaluate the potential usefulness of a PLR. To detect
differences in the zeropoint, different PLRs based on independently
determined distances are needed. We compare the $JH\Ks$ PLRs with
those of \citet{Chen16b}, which are based on open cluster distances
and {\sl Hippacos} parallaxes. In Fig. \ref{f1}, the cyan data points
represent the 55 CBs from \citet{Chen16b}, while the red dashed lines
are their best fits. No obvious zeropoint differences are
found. Statistically, the zeropoint differences between this paper and
\citet{Chen16b} are $\Delta J =-0.04\pm0.23$ mag, $\Delta H
=0.00\pm0.20$ mag, and $\Delta \Ks =0.00\pm0.21$ mag.

\subsection{Application of PLRs}

As distance indicators, the CB PLRs have thus been established and
calibrated based on the primary, direct distance determinations
afforded by the availability of TGAS parallaxes for 183 nearby CBs
(Sections 3.2 and 3.3). These PLRs can now be used as secondary
distance diagnostics to measure distances to individual CBs at greater
distances. For example, in the $W1$ band, the distance modulus for a
given CB could be determined by application of the equation ${\rm
  DM}=W1_{\rm max}-M_{W1{\rm max}}-A_{W1}$, where $M_{W1{\rm
    max}}=(-5.67 \pm0.17) \log P + (-0.15\pm0.08)$; $W1_{\rm max}$ and
$A_{W1}$ are the maximum apparent magnitude and the extinction in the
$W1$ band, respectively. The corresponding distance (modulus)
uncertainty propagating from the uncertainties in the PLRs is
equivalent to the 1$\sigma$ scatter in Table \ref{t2}, i.e., it is
0.16 mag for the $W1$ band. Therefore, we would expect a limiting
accuracy of 0.16 mag for individual CBs (statistical error). Sections
4 and 5 below are based on application of this principle, aimed at
determination of the distances to individual CBs, both in open
clusters and in the Galactic field.

\section{Distance Measurements to Open Clusters}
W UMa-type CBs are usually found in old open clusters, since their
ages span the 1--10 Gyr range. The distance accuracy of CBs is similar
to that for open clusters based on main-sequence fitting (typically
5\% statistical and 5\% systematic uncertainties). Based on
application of their PLRs, CBs can provide independent distances to
open clusters. \citet{Chen16b} listed 19 open clusters containing W
UMa-type CBs. We used the $G$ and $W1, W2$-band PLRs to estimate their
distances, since most CBs have multi-epoch magnitudes in these
bands. The mean-magnitude PLRs were adopted, since they are more
accurate for the faint CBs. As regards the extinction, the $E(B-V)$
values for the sample open clusters from literature were adopted, and
the $R_V=3.1$ extinction law \citep{Cardelli89} was used to estimate
$A_V$. For the $G$-band extinction, $A_G/A_V=0.87$ \citep{Jordi10} was
adopted. In the $W1$ and $W2$ filters, the extinction is about 0.05
$A_V$ \citep{Wang18}. We assume that the extinction affecting the
cluster CBs is similar to that pertaining to the host clusters
themselves.

\begin{figure}[h!]
\centering
\vspace{-0.0in}
\includegraphics[angle=0,width=160mm]{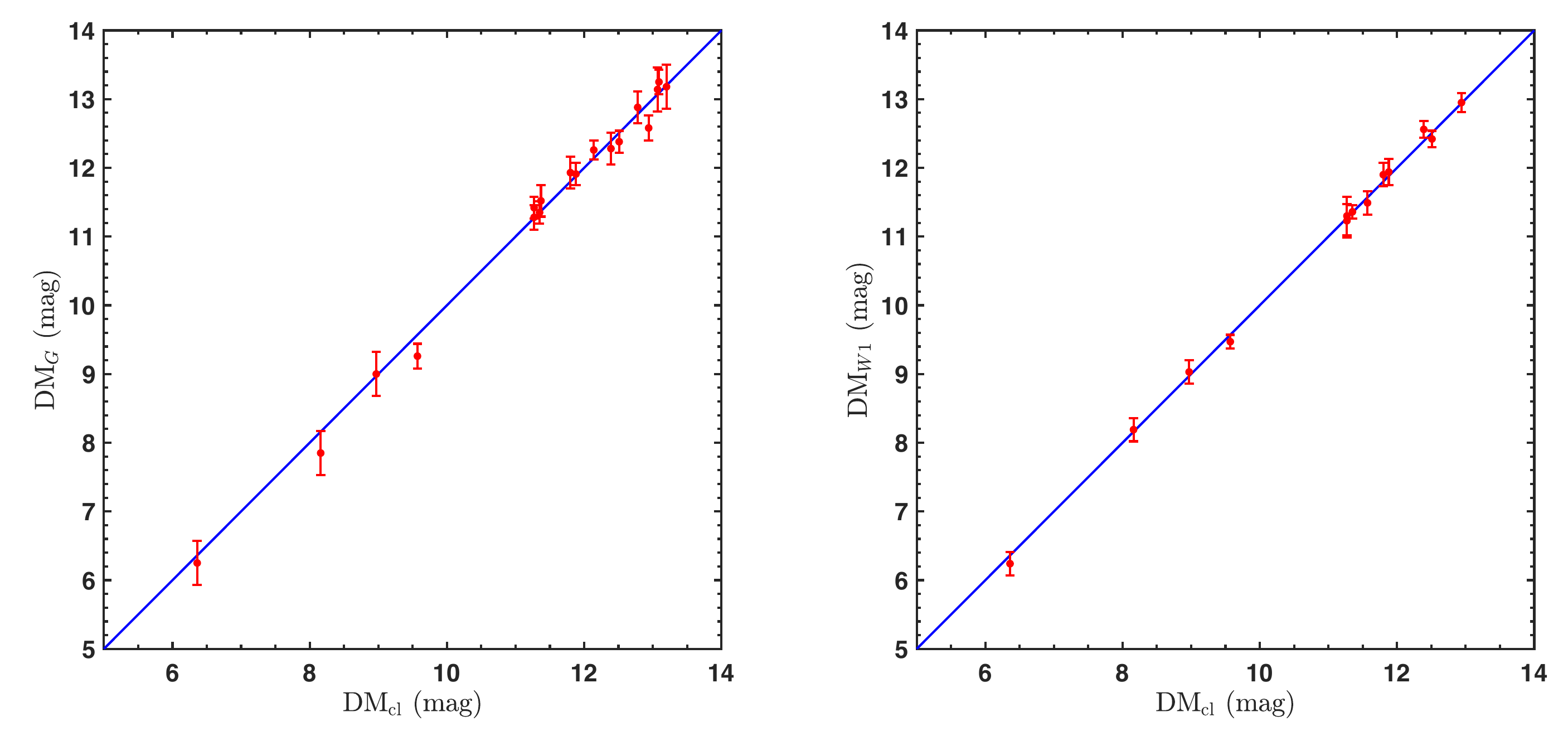}
\vspace{-0.0in}
\caption{\label{f3} Comparison of CB PLR distances with distances
  based on the open cluster method. The left-hand panel is based on
  the $G$-band PLR, while the right-hand panel shows the results for
  the $W1$ band.}
\end{figure}

The derived distance moduli are denoted as ${\rm DM}_G$, ${\rm
  DM}_{W1}$, and ${\rm DM}_{W2}$ in Table \ref{t3}. The error
pertaining to the distance modulus is the larger of the statistical
errors for the different CBs in a given cluster and the uncertainty
propagating from the PLR, $\sigma=\sigma_{\rm PLR}/\sqrt{N}$. Compared
with the distances determined using the open cluster method, the
differences are ${\rm DM}_G- {\rm DM_{cl}} = -0.02\pm0.16$ mag and
${\rm DM}_{W1}- {\rm DM_{cl}} = 0.00\pm0.08$ mag. The maximum
deviation of 1\% in these distance scales shows that both calibrations
are in good mutual agreement. In particular, the $W1$ distances are a
little better than the $G$-band distances in Fig. \ref{f3}. The two
clusters located more than their 1$\sigma$ uncertainties from the
one-to-one relation suggest that, at optical wavelengths, PLR
distances are sensitive to extinction and metallicity variations.

\begin{table}[h!]
\begin{center}
\caption{\label{t3}Comparison of Distance Moduli}
\vspace{0.15in}
\begin{tabular}{lcccccc}
\hline
 \hline 

 ID          &  ${\rm DM}_G$          &     ${\rm DM}_{W1}$    &   ${\rm DM}_{W2}$      &  $E(B-V)$ & $\rm{DM}_{OC}$ & Ref.$^a$\\ 
 & (mag) & (mag) & (mag) & (mag) & (mag) \\
\hline 

NGC 188      & $11.35\pm0.16$   & $11.36\pm0.10$   & $11.46\pm0.14$   &   0.09    & 11.35   &    1       \\ 
NGC 2682     & $9.26 \pm0.18$   & $9.47 \pm0.10$   & $9.47 \pm0.10$   &   0.01    & 9.57    &    2       \\ 
NGC 2158     & $12.88\pm0.23$   &                  &                  &   0.33    & 12.78   &    3       \\ 
NGC 1245     & $12.28\pm0.23$   & $12.56\pm0.12$   &                  &   0.25    & 12.39   &    3       \\ 
Berkeley 39  & $12.58\pm0.18$   & $12.95\pm0.14$   & $12.96\pm0.27$   &   0.17    & 12.94   &    4       \\ 
NGC 6791     & $13.25\pm0.18$   &                  &                  &   0.16    & 13.09   &    5       \\ 
NGC 6819     & $11.91\pm0.16$   & $11.94\pm0.19$   &                  &   0.14    & 11.88   &    5       \\ 
NGC 7789     & $11.42\pm0.16$   & $11.30\pm0.28$   & $11.39\pm0.17$   &   0.28    & 11.27   &    6       \\ 
NGC 2099     &                  & $11.49\pm0.17$   & $11.57\pm0.18$   &   0.23    & 11.57   &    7       \\ 
NGC 6705     & $11.52\pm0.23$   &                  &                  &   0.42    & 11.37   &    8       \\ 
Collinder 261& $12.26\pm0.14$   &                  &                  &   0.34    & 12.14   &    3       \\ 
Melotte 66   & $13.18\pm0.32$   &                  &                  &   0.16    & 13.20   &    9       \\ 
NGC 7142     & $11.93\pm0.23$   & $11.90\pm0.17$   & $11.95\pm0.18$   &   0.35    & 11.80   &   10       \\ 
NGC 6939     & $11.28\pm0.18$   & $11.23\pm0.24$   & $11.23\pm0.26$   &   0.31    & 11.27   &    3       \\ 
NGC 7044     & $12.38\pm0.16$   & $12.42\pm0.12$   & $12.53\pm0.18$   &   0.70    & 12.43   &   11       \\ 
NGC 2044     & $13.14\pm0.32$   &                  &                  &   0.08    & 13.07   &   12       \\ 
NGC 2184     & $ 9.00\pm0.32$   & $9.03 \pm0.17$   & $9.02 \pm0.18$   &   0.10    & 8.97    &    3       \\ 
Ruprecht 56  & $ 7.85\pm0.32$   & $8.19 \pm0.17$   & $8.20 \pm0.18$   &   0.12    & 8.16    &    3       \\ 
Praesepe     & $ 6.25\pm0.32$   & $6.22 \pm0.17$   & $6.22 \pm0.18$   &   0.01    & 6.36    &    3       \\ 

\hline
\end{tabular}
\tablenotetext{a}{(1) \citet{Chen16a}; (2) \citet{Geller15}; (3)
  \citet{Kharchenko16}; (4) \citet{Bragaglia12}; (5) \citet{Wu10}; (6)
  \citet{Wu07}; (7) \citet{Hartman08}; (8) \citet{Santos05}; (9)
  \citet{Kassis97}; (10) \citet{Straizys14}; (11) \citet{Sagar98};
  (12) \citet{Jacobson11}.}
\end{center}
\end{table}

\section{27,000 W UMa-type contact binaries with an 8\% distance accuracy}

CBs are not only found in open clusters, but they are also the numerous variable stars in the field. In this section, we collected a
sample of 55,603 CB candidates from the GCVS, ASAS, and Catalina
catalogs (see Section 2.2) and determined their distances based on
PLRs. With known distances and light curve solutions, the absolute
masses and radii can be estimated. These parameters are important to
constrain the evolution of CBs \citep{Stepien06, Yildiz13}. In
addition, with access to tens of thousands of CBs with reliable
distances, they can be used to constrain the zeropoints, in
anticipation of {\sl Gaia} DR 2.

\subsection{Fourier analysis to select high-probability W UMa-type contact binaries}

Before measuring the distances, contaminating objects must be
eliminated. As discussed in Section 2.2, CBs in the ASAS catalog are
distinguished from semi-detached eclipsing binaries, but they are
still mixed with other types of variables. CBs in the Catalina catalog
contain both semi-detached eclipsing binaries and other types of
variables. Although CB samples in the ASAS and Catalina catalog have
already been selected by other authors \citep{Pojmanski05, Drake14,
  Drake17}, we further refined the CB sample based on light curve
analysis. Note that our application of any selection criterion is
aimed at reducing contamination rather than identification of genuine
CBs, since the latter is impossible based on our current
data. Nevertheless this approach is sufficiently reliable for
statistical purposes. A fourth-order Fourier series decomposition,
$\displaystyle{f =a_0 + \sum_{i=1}^4a_i\cos(4\pi it/P+\phi_i)}$, was
used to analyze the $V$-band light curves. Here, $P$ is the (orbital)
period for eclipsing binaries and twice the (pulsation) period for
other types of variable stars. Compared with the Fourier analysis
conducted by \citet{Rucinski93}, phase information is introduced to
distinguish eclipsing from pulsating variables. Note that $a_1$ and
$a_2$ in this paper are equivalent to, respectively, the absolute
values of $a_2$ and $a_4$ of \citet{Rucinski93}. We used the
parameters $\log P$, $R_{21}=a_2/a_1$, and $\phi_{21}=\phi_2-2\phi_1$
to select W UMa-type CBs:

\begin{equation}
   \label{equation1}
 -0.60<\log P {\rm [d]}<-0.25;
 \end{equation}
 \begin{equation}
   \label{equation2}
 a_2<a_1(a_1+0.125);
 \end{equation}
  \begin{equation}
   \label{equation3}
 R_{21}<0.5-0.5|\phi_{21}-2\pi|.
\end{equation}

\begin{figure}[h!]
\centering
\hspace{0.0in}
\includegraphics[angle=0,width=180mm]{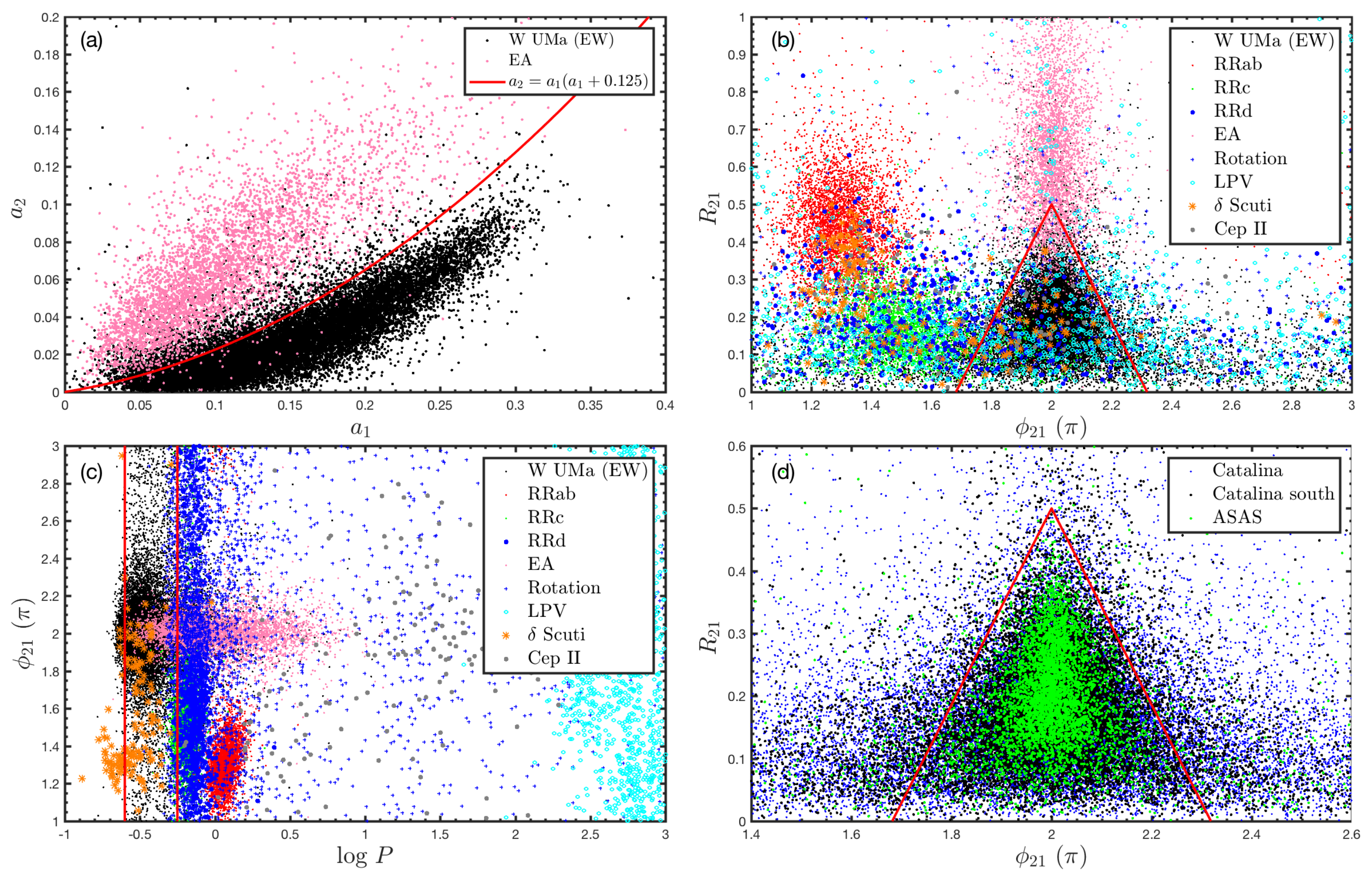}
\vspace{-0.0in}
\caption{\label{f4} (a) $a_2$ versus $a_1$ diagram. The black and pink
  data points are W UMa CBs (EW type) and detached eclipsing binaries
  (EA type), while the red line corresponds to
  $a_2=a_1(a_1+0.125)$. (b), (c) $R_{21}$ versus $\phi_{21}$ and
  $\phi_{21}$ versus $\log P$ [d] diagrams for different types of
  variable stars in the Catalina southern catalog. Black: W UMa-type
  CBs; red: RRab; green: RRc; blue: RRd; pink: EA-type eclipsing
  binaries; blue pluses: rotating variable stars; cyan circles:
  long-period variables (LPVs); orange asterisks: $\delta$ Scuti; grey
  dots: type II Cepheids; red lines: criteria defined by Equations (1)
  and (3). (d) $R_{21}$ versus $\phi_{21}$ diagram for W UMa-type CBs
  in the Catalina (blue), the Catalina southern (black), and the ASAS
  (green) catalogs. The red line are the same as those in panel (b).}
\end{figure}

The first criterion limits the period range, since PLRs would
overestimate the absolute magnitudes of CBs with periods outside of
this range. This period criterion will also reduce contamination by
variables with long periods and symmetric light curves; see Fig.
\ref{f4}c. The second criterion results from the theoretical work of
\citet{Rucinski93}, which aimed at distinguishing CBs from
semi-detached and detached eclipsing binaries. Unlike CBs, detached
binaries exhibit unequal minima in their light curves, which increases
the ratio $a_2/a_1$. In Fig. \ref{f4}a, the red line is the boundary
separating CBs (with EW-type light curves) and detached binaries
(EA-type light curves) based on the Catalina southern catalog. Note
that in that catalog, CBs and detached eclipsing binaries are mixed
with semi-detached eclipsing binaries, because semi-detached eclipsing
binaries are not explicitly selected.

The third criterion above is first introduced in this paper. It can be
used to exclude short-period variables with asymmetric light curves,
such as RRc and RRd Lyrae, and reduce contamination by $\delta$ Scuti
and rotating variable stars (see Fig. \ref{f4}b and \ref{f4}c). To
apply this criterion, the upper limit of $R_{21}=0.5$ is converted
from the second criterion using $a_1=0.75/2$ mag; 0.75 mag is usually
the maximum amplitude of CBs. The phase difference $\phi_{21}$, is
either 0 or 2$\pi$ in ideal conditions (we used 2$\pi$ throughout to
render an uninterrupted spread of CBs in Fig. \ref{f4}b and \ref{f4}c)
because of the symmetric light curves (regular eclipses) associated
with CBs. However, photometric uncertainties and possible surface
activity, or the presence of a third object, would cause $\phi_{21}$
to deviate from 2$\pi$. The extent of this deviation increases with
decreasing half main amplitude $a_1$ ($R_{21}$). We set the extent of
the deviation (slope) such that it covers the triangular distribution
of CBs in the three catalogs (see Fig. \ref{f4}d) and thus determined
this criterion.

Following application of our three criteria, the fractions of
high-probability W UMa-type CBs in the ASAS, Catalina southern, and
Catalina catalogs are 56\%, 64\%, and 49\%, respectively. As regards
the former two catalogs, most W UMa-type CBs candidates are excluded
based on their periods, while in the latter catalog, some candidates
are excluded based on their light curve shapes. This distinction is
reliable, since the Catalina catalog is deeper and contains a number
of low-amplitude CB candidates, which are easily mixed with other
types of variables.

\subsection{Distance determination}

In Section 4, we determined the distances to W UMa-type CBs in open
clusters and found that the mid-infrared $W1$ distances are more
accurate than the corresponding optical $G$-band distances. In
addition, to achieve the smallest scatter in the PLRs, $W1$ is our
first choice to determine distances. Unlike CBs in clusters, for field
CBs, the photometric errors and extinction must be carefully dealt
with. In terms of dealing with the effects of extinction, $W1$ is much
better than the $G$ band, since its extinction is only 5\% of
$A_V$. As regards the photometric errors, $\sigma_G$ is less than 0.02
mag, while for $W1$ the corresponding value is around 0.05
mag. Considering all these uncertainties, the $W1$ band could be used
to determine CB distances to a precision of 8\% (0.177 mag). The
derived distance moduli, ${\rm DM}_{W1}$, are listed in Table
\ref{t4}. To make sure that the distances to field (non-cluster) CBs
are accurate and unbiased, we estimated their distances based on the
combination of our multi-band PLRs. The $G_{\rm mean}$ and $JH\Ks$
magnitudes of 55,603 CBs were used to determine the distance moduli
${\rm DM}_G$ and ${\rm DM}_{JHKs}$. Combined with ${\rm DM}_{W1}$, the
average distance becomes ${\rm DM}=\sum({\rm
  DM}_i/\sigma_i^2)/\sum(1/\sigma_i^2)$, where the weight $\sigma_i$
is the error in the PLR in a given band $i$. The uncertainty
$\sigma_{\rm DM}$ is the larger of the DMs' standard deviation and the
external PLR uncertainty of 0.16 mag. The average distance moduli and
their uncertainties are listed in Table \ref{t4}. Given the unknown
extinction affecting individual measurements, we recommend that only
${\rm DM}_{W1}$ and the average distance moduli ${\rm DM}$ based on
application of multiple bands be used (see the discussion of
extinction below). Table \ref{t4} also contains information as to
whether the CBs obey the criteria discussed in Section 5.1
(`Selection') as well as their multi-band photometric magnitudes.

  We remind the reader that the CBs in our global sample are
  located in the solar neighborhood (ASAS) and at high Galactic
  latitudes ($|b|>20^\circ$) (Catalina Survey), where the interstellar
  extinction is very small, even in the $V$ band. To evaluate the
  extinction properties, the color excess $E(G-W1) = (G-W1)-(G-W1)_0$
  was determined for all CBs individually by calculating the
  difference between their observed and intrinsic colors. For our
  final sample of 27,318 CBs, the average color excess is
  $E(G-W1)=0.156\pm0.274$ mag, which corresponds to $A_V=0.19\pm0.33$
  mag for the extinction law adopted here. 

Our distances were determined in two ways, i.e., based on the $W1$
band alone and on a combination of the $GJHKW1$ bands. Both methods
are insensitive to the actual extinction. The former method is
affected by a relative extinction $A_{W1}=0.05A_V$, while for the
latter, $A_{GJHKW1}=0.1A_V$. In fact, even if we had not corrected
these 27,318 CBs for the effects of extinction, the impact on the
determination of the final distance modulus is $\sigma=0.010\pm0.016$
mag (0.4\%) for the $W1$-based distances and $\sigma=0.019\pm0.033$
mag (0.9\%) for the multi-band distances. These systematic and
statistical uncertainties are both much lower than the distance
uncertainty introduced by the scatter in the PLR (0.16 mag), which
implies that the impact of extinction variations is negligible. Our
extinction correction is aimed at excluding any global trends in the
extinction properties and thus at reducing the statistical
uncertainties. Although accurate extinction measurements for
individual CBs are not available, the global distribution of our
sample distances is expected to be improved compared with their
corresponding distances obtained without corrections applied.

The extinction equation introduced in Section 3.2, $A_V=\rho_0
\int_{0}^{d}{\exp(-r|\sin(b)|/H){\rm d}r}$, was adopted to estimate
the individual extinction values for all CBs in our sample. The
adopted distances were based on the $W1$ apparent distance moduli,
which are mostly reddening-free on account of this filter's long
wavelength. To verify the applicability of this equation, we
calculated the average corrected extinction value for our 27,318 CBs,
i.e., $A_V=0.194\pm0.080$ mag. This value is in full accordance with
the observed value $A_V=0.19\pm0.33$ mag, which means that we are
indeed justified in correcting for the average extinction trend using
this global equation. The $A_{W1}$ extinction values are included in
Table \ref{t4}; we do not provide the individual $A_V$ values because
of the larger intrinsic variations at this wavelengths.

\begin{table}[h!]
\begin{center}
\caption{\label{t4} Multi-band magnitudes and distance moduli for
  55,603 CBs.\tablenotemark{b}} \tiny
\vspace{0.1in}
\begin{tabular}{rrccccccccccc}
\hline \hline
R.A. (J2000) &  Dec. (J2000) & Selection & Period  & $A_{W1}$ & ${\rm DM}_{W1}$  & ${\rm DM}$ & $\sigma_{\rm DM}$   
& ${\rm DM}_G$   & ${\rm DM}_{JHKs}$  & Amp.  &  $\langle G \rangle$ &  $\langle W1 \rangle$\\ 
 ($^\circ$) & ($^\circ$)& y/n & [d] & (mag) & (mag) & (mag) & (mag) & (mag) & (mag) & (mag) & (mag)& (mag)\\
\hline
0.13125&  -8.78119& n &0.40419& 0.00477& 10.910& 10.884& 0.155& 10.766& 10.893& 0.120& 14.418& 13.163\\
0.15392&  41.46825& y &0.27463& 0.01257& 11.665& 11.760& 0.155& 12.091& 11.761& 0.730& 17.319& 14.904\\
0.15646&  39.05225& n &0.30691& 0.01143& 12.300&  0.000& 0.000& 14.065& 12.527& 0.230& 18.858& 15.256\\
0.26442&  39.96236& n &1.96701& 0.01188& 14.934&  0.000& 0.000& 17.758& 15.379& 0.140& 15.641& 13.188\\
0.27900&  12.10286& n &0.29732& 0.00587& 11.107&  0.000& 0.000&  0.000& 11.073& 0.110&  0.000& 14.138\\
0.29537&  40.08919& n &0.66070& 0.01151& 10.829&  0.000& 0.000& 11.794& 11.069& 0.090& 13.734& 11.844\\
0.42200&  42.18561& y &0.30860& 0.01311& 14.194&  0.000& 0.000& 13.521&  0.000& 0.540& 18.323& 17.138\\
0.42617&  37.83233& y &0.35954& 0.01087& 12.001& 11.965& 0.155& 11.843& 11.964& 0.140& 16.037& 14.556\\
0.51687&  28.94167& n &0.24460& 0.00665&  8.519&  8.466& 0.155&  8.456&  8.396& 0.180& 14.011& 12.045\\
0.52458&  30.10169& n &0.42380& 0.00829& 10.448& 10.361& 0.155& 10.096& 10.343& 0.110& 13.633& 12.584\\
0.52975&  42.80536& n &0.30700& 0.01271& 10.744& 10.729& 0.155& 10.833& 10.666& 0.160& 15.648& 13.701\\
0.56462& -17.31761& n &0.31421& 0.00458& 13.749& 13.527& 0.247& 13.539& 13.208& 0.460& 18.126& 16.639\\
0.57083&  24.47011& n &0.28586& 0.00735& 11.676& 11.672& 0.155& 11.665& 11.668& 0.190& 16.652& 14.808\\
0.61271&  43.57911& n &0.42859& 0.01313& 10.763& 10.651& 0.191& 10.191& 10.675& 0.090& 13.771& 12.876\\
0.61413&  41.16000& n &0.34632& 0.01247& 12.409& 12.233& 0.219& 11.766& 12.172& 0.240& 16.127& 15.060\\
0.72329&  31.67353& n &0.28397& 0.00824&  9.719&  9.669& 0.155&  9.575&  9.636& 0.070& 14.602& 12.869\\
0.75117&  27.01969& n &0.44364& 0.00771& 10.592& 10.633& 0.155& 10.922& 10.576& 0.120& 14.279& 12.612\\
0.93129&   7.50461& n &0.76545& 0.00551& 13.339&  0.000& 0.000& 15.253& 13.482& 0.120& 16.540& 13.976\\
0.94379&  18.23014& n &0.26965& 0.00647& 13.946&  0.000& 0.000& 13.728&  0.000& 0.420& 18.918& 17.225\\
1.20017&  -5.82894& n &0.26783& 0.00455&  8.498&  8.559& 0.155&  8.919&  8.501& 0.120& 14.100& 11.792\\
1.23063&  39.84725& y &0.30155& 0.01174& 13.339&  0.000& 0.000& 13.067&  0.000& 0.240& 17.931& 16.340\\
1.25750&  32.62619& n &0.25838& 0.00907& 13.458&  0.000& 0.000& 13.971&  0.000& 0.510& 19.364& 16.848\\
1.27450&  44.46792& y &0.26832& 0.01409& 11.255& 11.251& 0.155& 11.319& 11.220& 0.370& 16.659& 14.554\\
1.32796&  42.13956& y &0.29282& 0.01290& 11.845& 11.775& 0.155& 11.549& 11.766& 0.180& 16.543& 14.922\\
1.55963&  26.54439& n &0.22607& 0.00764& 10.944& 11.071& 0.186& 10.898& 11.321& 0.170& 16.764& 14.670\\
1.59554&  -7.90947& n &0.24828& 0.00478& 11.503& 11.511& 0.155& 11.797& 11.408& 0.200& 17.264& 14.989\\
1.62058&  40.84006& n &3.79518& 0.01223& 19.086&  0.000& 0.000& 21.537& 19.194& 0.430& 16.978& 15.676\\
1.63500& -15.49064& n &0.22938& 0.00443&  8.751&  8.836& 0.240&  9.426&  8.722& 0.250& 15.182& 12.437\\
1.63904&  32.10075& n &0.49330& 0.00890& 12.043& 12.043& 0.159& 11.687& 12.185& 0.080& 14.669& 13.795\\
1.67675&  31.25036& n &0.34206& 0.00799&  9.544&  9.466& 0.155&  9.298&  9.423& 0.110& 13.628& 12.222\\
1.72925&  40.20617& n &0.24104& 0.01027&  9.709&  9.707& 0.155&  9.855&  9.644& 0.600& 15.528& 13.276\\
1.74221&  39.14983& n &0.32143& 0.01137& 13.278& 13.120& 0.166& 12.850& 13.006& 0.210& 17.471& 16.117\\
1.76996&  42.52472& n &0.29570& 0.01242& 10.689& 10.738& 0.155& 10.670& 10.835& 0.180& 15.620& 13.740\\
1.80767&  43.44511& y &0.30798& 0.01093&  9.507&  9.531& 0.155&  9.564&  9.551& 0.290& 14.336& 12.454\\
1.80812& -13.70717& n &0.25488& 0.00463& 12.399& 12.424& 0.155& 12.688& 12.354& 0.340& 18.054& 15.819\\
1.81571&  25.65161& n &0.27605& 0.00750& 11.868& 11.737& 0.155& 11.494& 11.648& 0.190& 16.614& 15.089\\
1.84858& -16.35031& n &0.32684& 0.00453&  9.413&  9.523& 0.250& 10.137&  9.434& 0.180& 14.576& 12.203\\
1.92046&  44.47922& y &0.44992& 0.01456& 13.190& 13.071& 0.155& 12.732& 13.037& 0.270& 16.156& 15.181\\
2.04150&  38.87400& n &0.30941& 0.01123& 14.144&  0.000& 0.000& 13.591&  0.000& 0.300& 18.350& 17.079\\
2.08767&  42.85478& y &0.25457& 0.01316& 11.525& 11.606& 0.155& 11.955& 11.581& 0.530& 17.475& 14.956\\
2.13142&  31.81703& n &0.30145& 0.00875& 11.048& 11.241& 0.208& 11.603& 11.369& 0.120& 16.417& 14.047\\
2.61287&  43.25250& n &0.27390& 0.01244& 10.453& 10.381& 0.155& 10.292& 10.314& 0.630& 15.527& 13.698\\
2.84687& -17.54211& y &0.35594& 0.00449&  9.273&  9.210& 0.188&  8.748&  9.304& 0.330& 12.869& 11.847\\
2.85125&  -4.10583& n &0.25445& 0.00477&  9.142&  0.000& 0.000& 10.770&  9.248& 0.280& 16.145& 12.566\\
2.87083& -10.18636& n &0.86288& 0.00469& 14.933&  0.000& 0.000& 16.248& 15.344& 0.280& 17.075& 15.266\\
2.87783& -12.80408& y &0.37177& 0.00463& 12.070& 12.006& 0.155& 11.942& 11.942& 0.360& 15.903& 14.534\\
2.91308& -16.44475& n &0.43543& 0.00457& 11.789& 11.813& 0.155& 12.104& 11.731& 0.100& 15.476& 13.853\\
2.94029&  18.26264& n &0.29767& 0.00641& 11.057& 11.112& 0.155& 10.904& 11.271& 0.340& 15.724& 14.086\\
2.94071&  18.26256& n &0.29766& 0.00641& 11.057& 11.112& 0.155& 10.904& 11.271& 0.330& 15.724& 14.086\\
2.95271& -12.08269& n &0.25525& 0.00465& 11.242& 11.094& 0.230& 10.551& 11.102& 0.190& 15.913& 14.658\\

    ... & ... & ... & ... & ... & ... & ... & ... & ... & ...& ...& ...&...\\ 

\hline
\end{tabular}
\tablenotetext{b}{The entire table is available as online data. A
  portion is shown here for guidance regarding its form and content.}
\end{center}
\normalsize
\end{table}

\begin{figure}[h!]
\centering
\hspace{0.0in}
\includegraphics[angle=0,width=180mm]{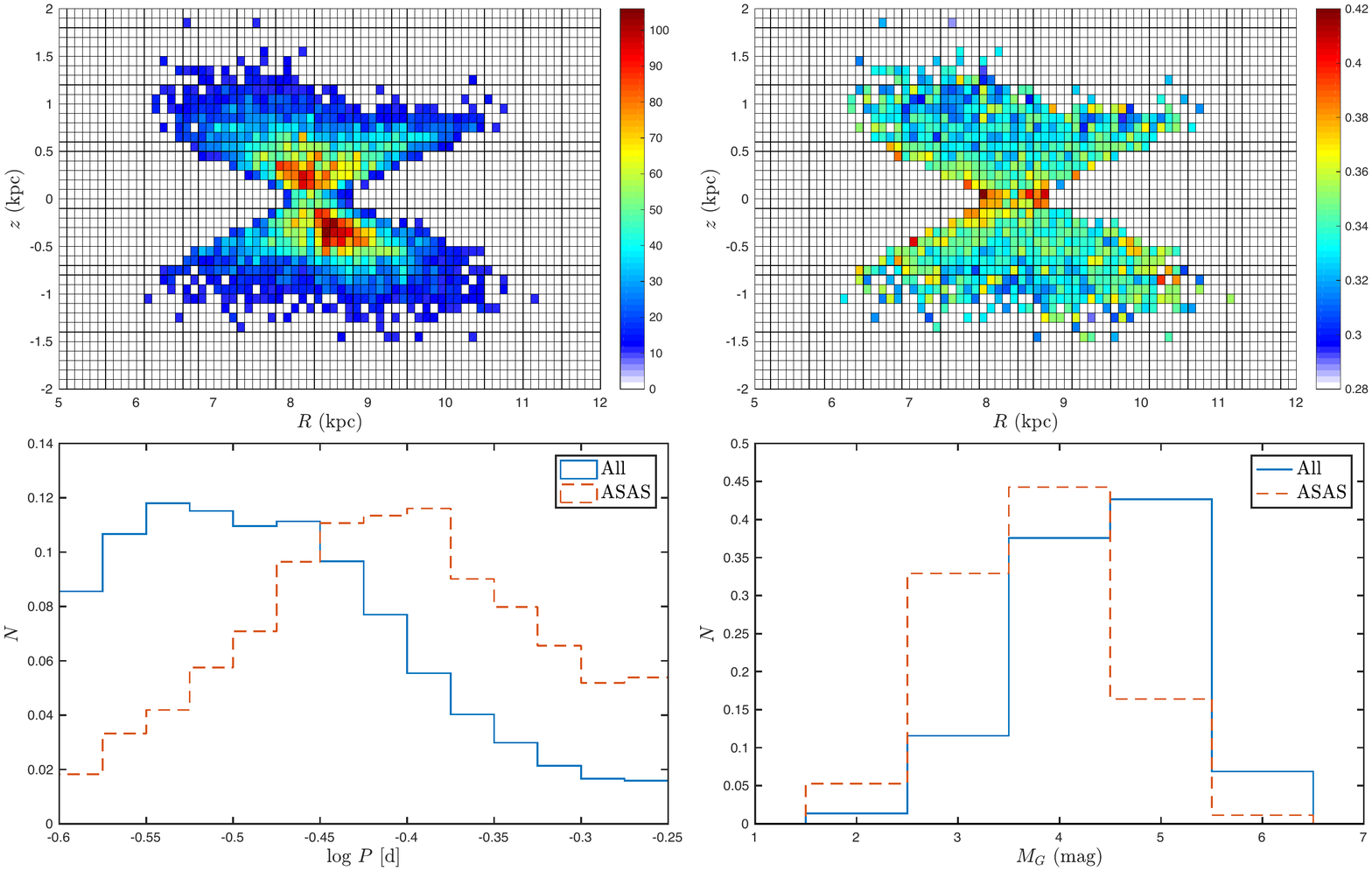}
\vspace{-0.0in}
\caption{\label{f5} (Top) Number distribution and orbital period
  spread of 27,318 W UMa-type CBs in the Galaxy. The size of the small
  boxes is $0.1 \times 0.1$ kpc$^2$. In color, the numbers and periods
  increase from blue to red. The bottom panels compare the normalized
  period and absolute $G$-band magnitude histograms of all CBs with
  those included in the ASAS catalog.}
\end{figure}

Finally, 27,318 of the 55,603 CB candidates obey the selection
criteria and have distance information based on multi-band
photometry. For these W UMa-type CBs, the average distance uncertainty
is 0.168 mag, which corresponds to a precision of 8\%. To study the
spatial distribution of these CBs in Galactic coordinates, the
distances were decomposed into a distance component along the Galactic
plane and a vertical component perpendicularly to the plane. The
vertical distances were then converted to Galactocentric distances by
assuming $R_0=8.3\pm0.2 \mbox{ (stat.)}\pm0.4$ (syst.) kpc
\citep{deGrijs16}. The numbers of CB candidates in each $0.1\times
0.1$ kpc$^2$ box were counted and are shown in the top left-hand panel
of Fig. \ref{f5}. We find that almost all CBs within 1 kpc have been
detected, except for those objects located in a region close to the
plane which was not covered by the Catalina survey. CBs are powerful
distance tracers out to 3 kpc.

An asymmetric distribution of CBs above and below the Galactic plane
is found. However, this result is not conclusive, since their velocity
information is not known. The average orbital periods were estimated
if the number of CBs in a given selection box exceeded 10. From the
period distribution in the top right-hand panel of Fig. \ref{f5}, it
follows that the longer-period CBs are located closer to the Galactic
plane. For W UMa-type CBs within $|z|<0.5$ kpc, the periods are
typically longer than 0.35 days. In the (vertically) outer regions,
the periods are on average shorter than 0.35 days. This is in
accordance with CBs in open clusters, where longer-period CBs are
found in younger open clusters \citep{Chen16b}.

The bottom panels of Fig. \ref{f5} show the distributions of the
periods and absolute magnitudes $M_G$ of all 27,318 W UMa-type CBs as
well as the equivalent measurements based on the ASAS catalog
only. The distribution of CBs in the ASAS catalog was discussed in
detail by \citet{Rucinski06}. Compared with the ASAS CBs, our full
sample is characterized by typically shorter periods and fainter
magnitudes. The main reason for this difference is that CBs in the
Galactic thick and thin disks are distributed differently. In the
young thin disk, long-period CBs encompass a larger fraction of the
overall sample. A second reason is that the shallow limiting magnitude
of ASAS ($V \sim 14$ mag) prevents detection of shorter-period CBs.

\section{Systematic uncertainties and improvements}

\subsection{PLR Systematics}

 In this section, we discuss the systematic uncertainties
  associated with the adoption of our set of PLRs for statistical
  analysis. Systematic effects include perturbations owing to tertiary
  companions, contamination, and metallicity corrections.

Many CBs have tertiary companions \citep{Chambliss92, Hendry98}, which
would render systems brighter. Only tertiary companions associated
with nearby CBs can be found on the basis of light travel time effects
in long-term O--C studies, subsequently resolved by spectral
studies. We evaluate to extent to which tertiary companions affect the
PLRs based on an analysis of nearby CBs. \citet{DAngelo06} performed a
spectroscopic search for faint ($V<10$ mag) tertiaries in CB
systems. They found that approximately 25\% of CBs have tertiaries
with higher luminosities than 0.9\% of those of the CB systems as a
whole. The mean and median luminosity ratios of the tertiaries in this
study are 11\% and 2\%, respectively. If these luminosity ratios are
representative for a large sample of CBs, the influence of tertiaries
on the resulting CB distances is less than 3\%. If additional distance
information is available that allows us to exclude outliers, the
effects will decrease to 0.5\%. A a case in point, distances based on
CBs in open clusters are statistically identical to those determined
using the clusters' intrinsic photometric properties (see Section 4).

Although CBs have characteristic light curves, they are nevertheless
easily mixed with other variables, particularly when their amplitudes
are low or for poor-quality light curves. In Fig. \ref{f4}c, CBs are
located close to Type-c RR Lyrae (RRc) and rotating stars. Compared
with RRab Lyrae, the light curves of RRc Lyrae are more symmetric and
have lower amplitudes. To distinguish RRc Lyrae from CBs,
\citet{Drake14} adopted three criteria based on the objects' periods,
amplitudes, and color differences. They concluded that the fraction of
RRc Lyrae that are misidentified as CBs is only on the order of
1\%. RRc Lyrae are some 2 magnitudes brighter than CBs for different
periods. Therefore, contamination leads to larger CB distances by on
the order of 1\% (0.02 mag). 

Other possible contaminants are rotating BY Dra variables and RS CVn
variables. The light variations of BY Dra variables are caused by
strong surface activity and rotation. However, they have a typical
rotational speed of 3--5 km s$^{-1}$, which implies that the majority
of these stars have periods longer than a day. Based on the GCVS
sample, the contamination fraction is less than 1\% in terms of the
objects' periods. Combined with the differences between BY Dra stars
and CBs in the shapes of their light curves (i.e., $\phi_{21}$ in
Fig. \ref{f4}), the associated contamination of our CB sample is less
than 0.2\%. The absolute magnitudes of BY Dra stars
\citep{Strassmeier93} are similar to or a little fainter than those of
CBs, so the effect of BY Dra contamination on the CB distances is
negligible. RS CVn variables are close binaries with strong surface
activity. They are usually found in relatively longer-period detached
and semi-detached binaries. The period overlap of GCVS-based RS CVn
and CB samples is approximately 3\%. By adding a light-curve selection
criterion, this fraction decreases to 0.6\%. Since RS CVn stars have
similar luminosities as CBs, the effect of this type of contamination
on CB distances is also negligible.

Finally, metallicity effects on the CB PLRs are not well studied
because of the prevailing limited sample sizes. Based on
\citep{Chen16b}, however, CBs will become brighter by about 0.2 mag in
the NIR when the metallicity decreases by 1 dex in [Fe/H]. No
systematic deviations are apparent when we determine the distances to
CBs using the mean solar abundance, but a correction is needed for
metal-poor or metal-rich environments.

\subsection{Improvement of the PLRs}

 The accuracy of the CB PLRs is 7\%, which is expected to be
  further improved based on more detailed studies. In this section, we
  discuss possible ways to improve the distance accuracy achievable.

The scatter in the CB PLR is composed of the scatter in the
mass--luminosity relation of the primary star, the
temperature--luminosity relation, the scatter in the mass ratio $q$,
the orbital inclination $i$, and the fill-out factor $f$. Detection of
discontinuous or nonlinear components of this scatter is helpful to
establish a more accurate PLR. The primary stars of CBs are similar to
main-sequence star; the latter obey a widely accepted mass--luminosity
relation. The scatter in the temperature--luminosity relation
decreases significantly from the optical to the infrared regime. 

We focus on the latter three contributions to the scatter here. The
mass ratio $q$ determines the size and area of the system's Roche
lobe. Based on the approximate radius of the Roche lobe
\citep{Eggleton83,Yakut05}, the areas of the inner and outer Roche
lobes increase significantly for $q<0.2$. The actual area for CBs is a
function $S(q,f)$, which covers a region between the inner and outer
Roche lobes. The fill-out factor $f$ represents the degree by which
the inner Roche lobe is exceeded; overcontact binaries have positive
values. For the same mass ratio, CBs with larger fill-out factors are
brighter. Since CBs are tidally distorted, their polar radii are
usually smaller than the radii on the side and back, which leads to
high luminosities for relatively low orbital inclinations.

\begin{figure}[h!]
\centering
\hspace{0.0in}
\includegraphics[angle=0,width=160mm]{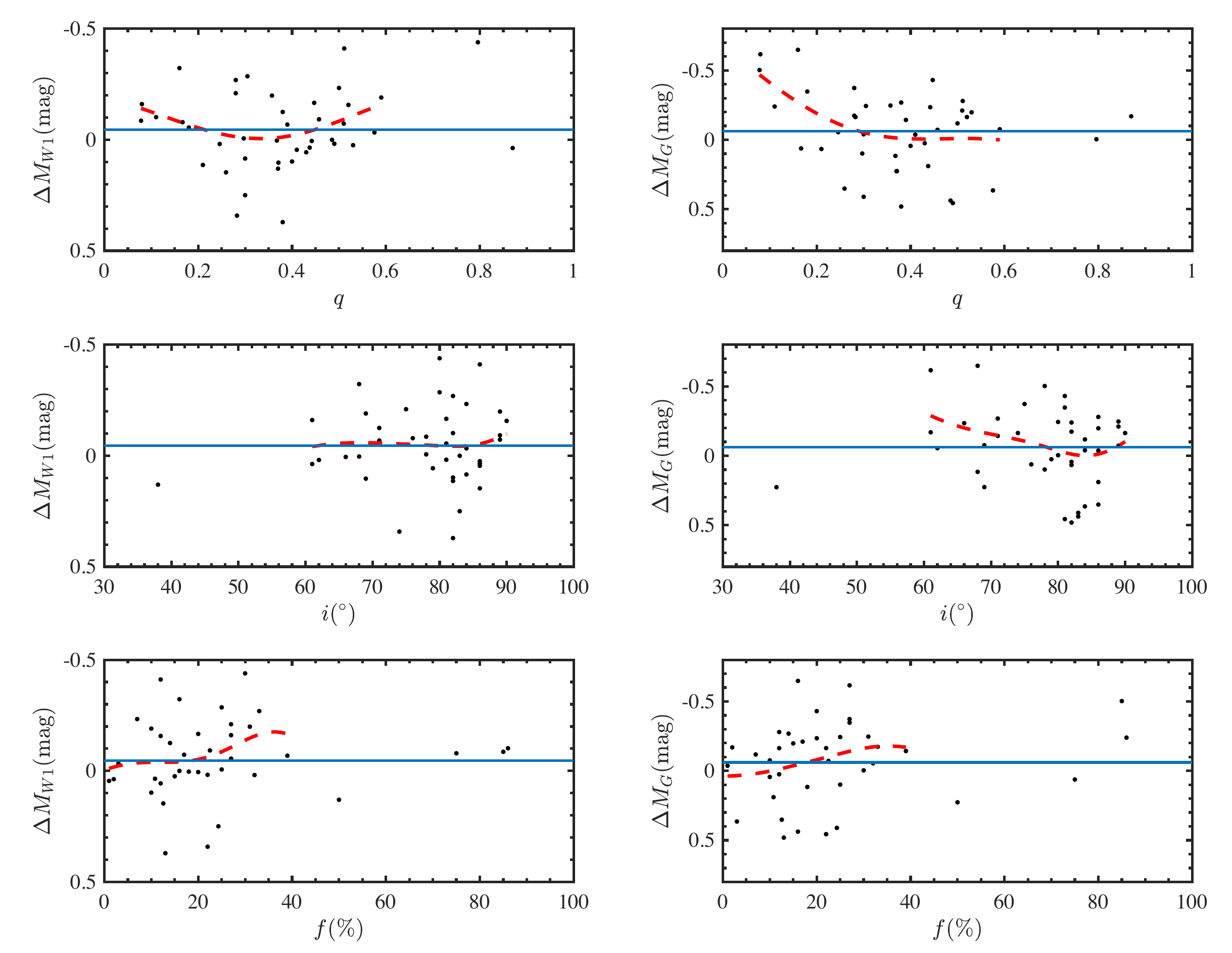}
\vspace{-0.0in}
\caption{\label{f6} Magnitude deviations as a function of (top) mass
  ratio $q$, (middle) orbital inclination $i$, and (bottom) fill-out
  factor $f$. Black dots denote our 41 CBs, while the blue lines
  represent the mean magnitude deviation. The red dashed lines are the
  best-fitting line based on application of a local linear kernel
  smoothing regression method.}
\end{figure}

We performed an observational test based on 41 of our 183 CBs using
these three bits of information. We estimated the deviations of the
PLRs, e.g., $\Delta M_{W1}=M_{W1}-M_{W1(\rm{PLR})}$. The $W1$ and $G$
bands were used to represent infrared and optical filters,
respectively. A local linear kernel smoothing regression method was
adopted to determine the nonlinear relation among $\Delta M_{W1},
\Delta M_{G}$, and $q,i,f$. The data points were divided into five
bins along the $x$ axis; we ignored any bin containing fewer than 2
data points. In Fig. \ref{f6}, the blue and red dashed lines are the
mean magnitude deviation and the local best-fitting line,
respectively. Comparing the two lines, local nonlinearities can be
found. In the $q$ panels (top), an obvious luminosity excess is found
in both the optical and the infrared for $q<0.2$. In the $i$ panels
(middle), a luminosity excess is found in the optical but it is not
see in the infrared for $60^\circ<i<72^\circ$. In the $f$ panels
(bottom), gradually increasing trends with $f$ are visible. In
addition, the excesses found for the global mean magnitudes $\Delta
M_{W1}=-0.045$ mag and $\Delta M_{G}=-0.061$ mag for our 41
overcontact binaries imply that overcontact binaries are
systematically brighter than near-contact binaries. In conclusion, CBs
with $q<0.2$ indeed deviate from the bulk PLRs; CBs with
$60^\circ<i<72^\circ$ may also deviate from the PLRs, although not
obviously so; the brightness of CBs increases as the fill-out factor
increases. These results should be better constrained not only by
employing accurate distances, but also by including more CBs with
homogeneous light curve solutions.

As regards improvement based on improved external conditions, better
and independent distances are expected to better constrain both the
zero points of and the scatter in the PLRs. Compared with Cepheids and
RR Lyrea, the PLRs of CBs are more difficult to constrain because of
the small sample sizes currently available. {\sl Gaia} Data release 2
will provide better parallaxes (0.07 mas precision) for more CBs
within 1 kpc of the Sun. Novel studies of CBs in open and globular
clusters are also expected to pick up, since clusters provide not only
independent distances, but also age and chemical abundance
information.
\section{Conclusions}
In this paper, 183 nearby W UMa-type CBs with accurate TGAS parallaxes
were used to determine the twelve-band $GBVRIJH\Ks W1W2W3W4$ PLRs. Not
only maximum-magnitude PLRs, but also mean-magnitude PLRs were
presented. The $1\sigma$ PLR dispersions decrease from optical to
mid-infrared wavelengths, with the lowest dispersion of 0.16 mag found
in the $W1$ band. Combining the $G$ and $W1$ bands, the PLCR exhibits
a scatter 0.16 mag, which means that W UMa-type CBs can anchor
distances to a 7\% accuracy. Since W UMa-type CBs are one of the most
numerous variable stars in the solar neighborhood, they could be
important distance indicators second only to classical Cepheids. No
obvious zeropoint differences are found for the near-infrared PLRs
between this paper and \citet{Chen16b}. The PLR of early-type CBs will
be better studied in the future based on larger samples with
independent distance estimates. Since OB-type CBs are brighter than
classical Cepheids, it is important to know the distances out to which
early-type CBs can be traced.

Applying the PLRs to W UMa-type CBs in 19 OCs, the CBs' PLR distances
agree well with the corresponding distances based on the open cluster
method. In particular, $W1$-band distances are better than their
$G$-band counterparts, since the former are insensitive to extinction
and metallicity variations. A catalog of 55,603 CBs candidates has
been compiled. Fourier decomposition was used to distinguish W
UMa-type CBs and other types of variable stars. To reduce the
uncertainties, distances based on a combination of optical, near-, and
mid-infrared distances were determined. Ultimately, 27,318 of the
55,603 CB candidates are high-probability W UMa-type CBs for which we
can achieve an 8\% distance accuracy. This is the largest sample with
accurate distances in the local volume, within a radius of 3 kpc. It
is useful to calibrate the zeropoints of other distance tracers and
constrain the absolute parameters of W UMa-type CBs.

We study the spatial distribution of the 27,318 W UMa-type CBs. A
trend that long-period W UMa-type CBs are found close to the Galactic
plane is found. This is in accordance with the properties of W
UMa-type CBs in open clusters. Compared with W UMa-type CBs in the
ASAS catalog, the full sample contains more short-period and faint W
UMa-type CBs. This suggests that in different environments, the
luminosity function of W UMa-type CBs may be different. Increased
numbers of W UMa-type CBs in the Galactic plane are expected to
facilitate better studies of the formation and evolution of W UMa-type
CBs. 

Possible systematic uncertainties introduced by tertiary
  companions, contamination, and metallicity differences are discussed
  in the context of applying the resulting PLRs to measure
  distances. The relations between the PLRs on the one hand and $q, i,
  f$ on the other are are discussed. The availability of better and
  more independent distances is expected to further constrain both the
  zero points of and the scatter in the CB PLRs.

\acknowledgments{We thank the anonymous referees from their helpful comments. We are grateful for research support from the
  National Natural Science Foundation of China through grants
  U1631102, 11373010, and 11633005, from the Initiative Postdocs
  Support Program (No. BX201600002), and from the China Postdoctoral
  Science Foundation (grant 2017M610998). This work was also supported
  by the National Key Research and Development Program of China
  through grant 2017YFA0402702.}

\end{document}